\PassOptionsToPackage{pdfpagelabels=false}{hyperref}
\documentclass[fleqn,usenatbib]{mnras}

\usepackage{newtxtext,newtxmath}

\usepackage[T1]{fontenc}
\usepackage{aecompl}

\usepackage{graphicx}	
\usepackage{amsmath}	
\usepackage{amssymb}	
\usepackage{multirow}
\usepackage{color}
\usepackage{ulem}

\title[Large elliptical disk in TDE ASASSN-14li]{A large accretion disk of extreme eccentricity in the
TDE ASASSN-14\lowercase{li}}

\author[Cao et al.]{R. Cao,$^{1}$
F.K. Liu,$^{1,2}$\thanks{E-mail: fkliu@pku.edu.cn (FKL)}
Z.Q. Zhou,$^{1}$
S. Komossa,$^{3}$
L.C. Ho$^{2,1}$
\\
$^{1}$Department of Astronomy, School of Physics, Peking University, Beijing 100871, China\\
$^{2}$Kavli Institute for Astronomy and Astrophysics, Peking University, Beijing 100871, China\\
$^{3}$Max-Planck-Institut f\"ur Radioastronomie, Auf dem H\"ugel 69, 53121 Bonn, Germany
}
\date{Accepted XXX. Received YYY; in original form ZZZ}

\pubyear{2017}

\begin{document}
\label{firstpage}
\pagerange{\pageref{firstpage}--\pageref{lastpage}}
\maketitle

\begin{abstract}
In the canonical model for tidal disruption events (TDEs), the stellar debris circularizes quickly to form an
accretion disk of size about twice the orbital pericenter of the star. Most TDEs and candidates discovered
in the optical/UV have broad optical emission lines with complex and diverse profiles of puzzling origin.
Liu et al. recently developed a relativistic elliptical disk model of constant eccentricity in radius
for the broad optical emission lines of TDEs and well reproduced the double-peaked line profiles of the TDE candidate PTF09djl with a large and extremely eccentric accretion disk. In this paper, we show
that the optical emission lines of the TDE ASASSN-14li with radically different profiles are well modelled
with the relativistic elliptical disk model, too. The accretion disk of ASASSN-14li has an eccentricity 0.97
and semimajor axis of 847 times the Schwarzschild radius ($r_{\rm S}$) of the black hole (BH). It forms as
the consequence of tidal disruption of a star passing by a massive BH with orbital pericenter $25 
r_{\rm S}$. The optical emission lines of ASASSN-14li are powered by an extended X-ray source of flat
radial distribution overlapping the bulk of the accretion disk and the single-peaked asymmetric line
profiles are mainly due to the orbital motion of the emitting matter within the disk plane of 
inclination about $26\degr$ and of pericenter orientation closely toward the observer. 
Our results suggest that modelling the complex line profiles is powerful in probing the structures of 
accretion disks and coronal X-ray sources in TDEs. 
\end{abstract}

\begin{keywords}
accretion disk --- black hole physics ---  galaxies: active --- line: profiles --- quasars: supermassive black holes
\end{keywords}



\section{Introduction} \label{sec:intro}

A star is tidally disrupted when it passes too close to a supermassive black hole (SMBH) in a
galactic nucleus \citep{hil75,ree88,eva89,lod09,gui13,hay13}. About half the stellar debris
after disruption becomes bound to the SMBH and returns to the orbital pericenter of the star.
Hydrodynamic simulations of stellar tidal disruptions suggest that the returned bound stellar debris 
self-interacts because of relativistic apsidal precession of the debris orbits and circularizes on
a timescale of a few times the return time of the most bound stellar debris to form an accretion disk
 \citep{ree88,koc94,hay13,gui14,pir15,shi15,bon16,hay16,sad16}. In the canonical scenario, the 
accretion disk is nearly circular and has size about twice the orbital pericenter of the star 
\citep[but see][]{shi15}. In this scenario, the light curves in the X-ray band are expected to 
follow the $t^{-5/3}$ fallback rate of stellar debris  but show some strong deviations in the optical/UV 
band. Stellar tidal disruption events (TDEs) discovered in the X-rays are consistent with these expectations,
but those in the optical/UV band are not \citep[][for a recent review]{kom99,gez06,van11,kom16}.

Most known TDEs and candidates discovered in the optical/UV have strong broad optical emission lines
with complex, asymmetric, and diverse profiles \citep{kom08,wan12,gez12,arc14,hol14,hol16a,hol16b,
cen16,fre16} and display large diversity in the relative strength of the \ion{He}{II} and Balmer emission lines
\citep{arc14,fre16}. Among them,  some have abnormally strong \ion{He}{II} emission lines while some have
weak or absent He emission lines in the optical spectra \citep{kom08,gez12,wan12,arc14,gas14,gez15,fre16,
hol16a,hol16b}, whose nature and origin are not yet fully understood. Some recent models for the spectra
of TDEs focus on the intensity ratios and widths of the optical emission lines \citep{gez12,gui14,gas14,
bog14,str15,koc16,kro16,met16,rot16}. Even tighter constraints  on the origin of the emission lines
may come from a more detailed investigation of the line profile shape and its variabilities.

Based on the hydrodynamic simulations for stellar tidal disruptions given in the literature, we recently
developed a relativistic elliptical accretion disk model with constant eccentricity in radius for the 
broad optical emission lines of TDEs and
well reproduced the complex substructures of the extremely broad double-peaked H$\alpha$ emission
line of the TDE candidate PTF09djl \citep{liu17}. \citet{liu17} showed that the peculiar double-peaked
profiles of the H$\alpha$ emission line of  PTF09djl with one peak at the line rest wavelength and
the other redshifted to about $3.5\times 10^4 \, {\rm km \; s^{-1}}$ can be well modeled with an elliptical
disk of a semimajor axis of 340$r_{\rm S}$ with $r_{\rm S}$ the BH Schwarzschild radius, large inclination 
$88\degr$, and extreme constant eccentricity $0.966$ in radius. A large elliptical accretion disk of 
constant eccentricity in radius suggests a fast accretion of matter and slow circularization of stellar debris,
inconsistent with the canonical model for TDEs \citep{ree88} but well
consistent with the physical arguments for efficient viscous stress \citep{svi17} and  observational
constraints on the disk viscous time of a very small fraction of the return time of the most bound stellar 
debris \citep{moc18}. 

In this paper, we model the profile of the broad H$\alpha$ emission line of the TDE ASASSN-14li with
the relativistic elliptical disk model. The H$\alpha$ emission line of the TDE ASASSN-14li has a single-peaked
profile radically different from the double-peaked profiles of the broad H$\alpha$ emission line of
PTF09djl. We show that the complex and asymmetric substructures of the H$\alpha$ line profiles of the
spectra of ASASSN-14li are also well reproduced with the elliptical relativistic disk model. The accretion disk
of TDE ASASSN-14li has an extreme eccentricity $0.970$ and large semimajor axis $847.0 
r_{\rm S}$ and inclines with respect to the line of sight (LOS) by an angle about $26\degr$. The 
single-peaked asymmetric line profiles are mainly due to the orbital motion of the emitting matter within 
the disk plane of small inclination and pericenter orientation closely toward the observer. 
The optical emission lines are powered by an extended X-ray source of flat  radial distribution 
of radial extent about 1668.5$r_{\rm S}$, or the entire elliptical accretion disk. The
enhanced He emission lines in the optical spectra and low disk inclination of ASASSN-14li are
consistent with the scenario that the strong He emission lines are expected to form in an ionized
optically thick accretion disk and the viewing angle effects of disk inclinations drive the diversity
of the relative He emission lines in TDEs.

The paper is organized as follow. We analyze the spectra of ASASSN-14li in Section~\ref{sec:data}. The
formation of accretion disk in TDEs is presented in Section~\ref{ssec:disk}, and the disk model for the broad
emission lines is presented in Section~\ref{sec:model}. The disk modeling of the broad H$\alpha$ profiles
is given in Section~\ref{spec:fit}. In Section~\ref{sec:disk}, we give the inferred parameters of the elliptical
accretion disk.  Implications of our results are discussed in Section~\ref{sec:dis}.

\section{The spectral data of ASASSN-\lowercase{14li}}\label{sec:data}

\begin{figure}
\begin{center}
\includegraphics[width=0.9\columnwidth]{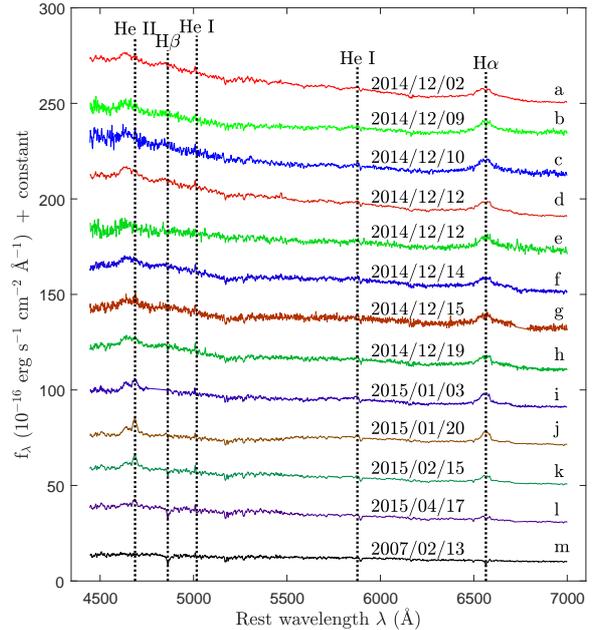}
\caption{The optical spectra of ASASSN-14li (a-l), corrected for Galactic extinction, arranged in a time
sequence from top to bottom and shifted vertically by arbitrary constants for clarity; the UT dates
(year/month/day) of the observations are given above each spectrum. The spectrum of the host galaxy
PGC 043234 is given at the bottom (m). Optical emission lines in the transient spectra, including Balmer
lines, \ion{He}{I}$\lambda5016$, \ion{He}{I}$\lambda5876$ and \ion{He}{II}$\lambda4686$, are
prominent and broad; they are identified with vertical dashed lines.
\label{fig:spec}
}
\end{center}
\end{figure}

The TDE ASASSN-14li was discovered in the redshift $z=0.0206$ galaxy PGC 043234 (VII Zw 211) on 2014
November 22, and 22 follow-up optical spectra were obtained between 2014 December 2 and 2016 
April 4  \citep{hol16a,bro17}. An optical spectrum of the host galaxy PGC 043234 is available from the Sloan 
Digital Sky Survey Data Release 9 \citep{ahn12}. The fluxes of the spectra were scaled so that the synthetic 
$r$-band magnitudes from the spectra match the contemporaneous $r$-band photometry \citep{hol16a}, 
although our results depend only on the line profiles and not on the absolute fluxes of the emission lines. 
We corrected the spectra for a Galactic extinction of $E(B-V) = 0.022 $ mag \citep{car89,sch11}. We use
stellar absorption lines to determine the systemic redshift of the host. The spectra of ASASSN-14li  and the
spectrum of host galaxy PGC 043234 are shown in Fig.~\ref{fig:spec}. The Balmer, \ion{He}{I}$\lambda
5016$, \ion{He}{I}$\lambda5876$, and \ion{He}{II}$\lambda4686$ lines except the possible 
\ion{He}{I}$\lambda6678$ are broad and prominent. In
Fig.~\ref{fig:spec} and the rest of the paper, we do not include the three spectra from the MDM 
Observatory Hiltner 2.4 telescope obtained  with the Ohio State Multi-Object Spectrograph with 
wavelength range 4200--6800~\AA~\citep{hol16a}, because the wavelength limit is too close to 
that of redshifted H$\alpha$. Nor are the spectra of 2015 February 4, 2015 May 19, 2015 December 09, 
2016 February 8, and 2016 April 4 included, whose signal-to-noise ratios are too low or the emission 
lines are to weak, with almost absent wings, to give a useful result.

\begin{figure}
\begin{center}
\includegraphics[width=0.9\columnwidth]{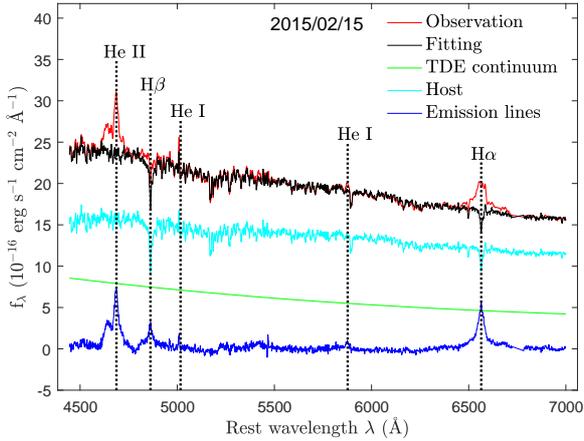}
\caption{Sample spectral decomposition of one of the spectra of ASASSN-14li (2015 February 15). After
correcting for Galactic extinction, the spectrum was fit with (1) a featureless TDE continuum, modeled with
a third-order polynomial (green),  and (2) host galaxy starlight (2007 February 13; light blue).  The pure
emission-line residual spectrum is shown in blue.  Emission lines, including Balmer lines,
\ion{He}{I}$\lambda5016$, \ion{He}{I}$\lambda5876$ and \ion{He}{II}$\lambda4686$, are identified
with vertical dashed lines.
\label{fig:exa}}
\end{center}
\end{figure}

To obtain a reliable measurement of the profiles of the broad emission lines for each epoch, we need to
subtract the continuum, which contains contributions from a featureless component from the TDE and a
component from the host galaxy starlight. The continuum of the TDE is modeled by fitting to several line-free
regions of the TDE spectrum with a third-order polynomial\footnote{Different procedures of continuum 
fitting with higher order polynomials were tested, and produce the same line results.}, plus a scaled host 
spectrum. The host spectrum is scaled by matching the regions with strong stellar absorption features and
without broad emission lines. Because the spectra of ASASSN-14li and the host galaxy were obtained using 
several telescopes with different spectral resolutions, we convolve all the spectra to match the lowest 
resolution of the spectra, $R\sim 8$~\AA. Fig.~\ref{fig:exa} illustrates our spectral decomposition 
procedure for one of the spectra.

\begin{figure}
\begin{center}
\includegraphics[width=0.9\columnwidth]{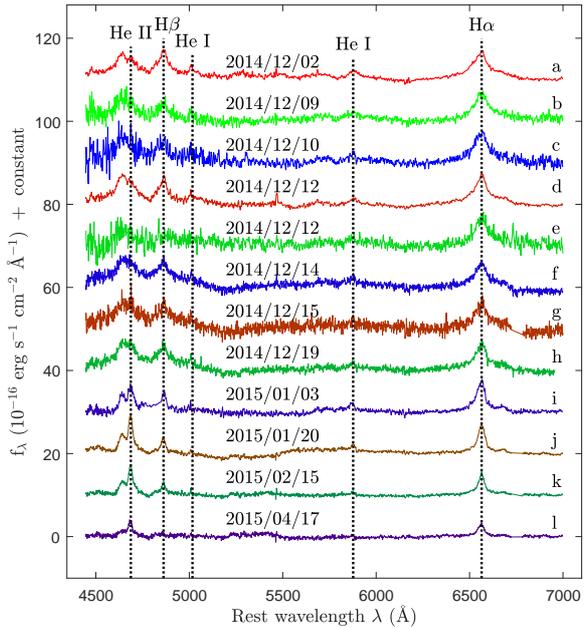}
\caption{Emission-line spectra of ASASSN-14li (a-l). The spectra are obtained after Galactic extinction
correction and subtraction of the continuum, globally modeled with a featureless TDE component and
host galaxy starlight component (see Fig. 2). They are arranged in a time sequence from top to bottom,
and shifted vertically by arbitrary constants for clarity; the UT dates (year/month/day) of the observations
are given above each spectrum. Broad emission lines are identified with vertical dashed lines.
\label{fig:line}}
\end{center}
\end{figure}

Fig.~\ref{fig:line} gives the 12 emission-line spectra of ASASSN-14li after continuum and host galaxy
starlight subtraction. H$\alpha$ is strong and broad, with full width at half maximum (FWHM) between
$1700$--$4500\, {\rm km \; s^{-1}}$ at all epochs. Other broad Balmer lines and \ion{He}{I}$\lambda5016$, 
\ion{He}{I}$\lambda5876$, and \ion{He}{II}$\lambda4686$ are also present, as initially reported 
by \citet{hol16a}.  H$\alpha$ is the strongest and least blended feature in the spectrum, and the continuum
is well-defined and relatively flat near H$\alpha$, minimizing uncertainties due to continuum subtraction.  The
other emission lines are blended with neighboring features and are more sensitive to the details of continuum
modeling. Therefore, we base our analysis and conclusions on the results of modeling the asymmetric and
complex profiles of the H$\alpha$ emission line.


\section{Modeling the broad emission lines of ASASSN-\lowercase{14li}}

\citet{liu17} recently set up a relativistic elliptical disk model for the broad optical emission
lines of TDEs. In this section, we present the disk model in some more detail and apply it to model the
spectra of ASASSN-14li.

\subsection{The accretion disk in TDEs}\label{ssec:disk}

A star is tidally disrupted by a SMBH when its orbital pericenter $r_{\rm p}$ is less than the tidal
disruption radius
\begin{equation}
r_{\rm t} = R_* \left(M_{\rm BH} / M_*\right)^{1/3} \simeq 23.54 r_* m_*^{-1/3} M_6^{-2/3} r_{\rm S},
\end{equation}
where $R_*= r_* {\rm R}_\odot$ and $M_*=m_* {\rm M}_\odot$ are, respectively, the star's radius and 
mass, and $M_{\rm BH} = 10^6  M_{\rm 6} {\rm M}_\odot$ is the  mass of the BH. When the bound stellar 
debris stream
returns to the pericenter, it is compressed and a nozzle shock forms at pericenter. The hydrodynamic
simulations of stellar tidal disruption indicate that the nozzle shock at the pericenter is weak and the
stellar debris stream is circularized due to the interaction of the post-pericenter outflowing and newly
inflowing streams because of relativistic apsidal precession \citep{eva89,koc94,hay13,dai15,shi15,pir15,
bon16,hay16}. The orbit of the most-bound stellar debris precesses by the instantaneous de Sitter
precession angle
\begin{equation}
    \Omega \simeq {6\upi G M_{\rm BH} \over c^2 a_{\rm mb} (1-e_{\rm mb}^2)}  \simeq
    {6\upi G M_{\rm BH} \over c^2 r_{\rm p} (1+ e_{\rm mb}) }
    \simeq {3\upi \over 2} {r_{\rm S} \over r_{\rm p} } = {3\upi \over 2} {1 \over x_{\rm p}}
    \label{eq:omega}
\end{equation}
with $ x_{\rm p}  = r_{\rm p} / r_{\rm S}$, where $e_{\rm mb} = 1 - \delta$ with $\delta \simeq 8.49
\times 10^{-4} r_*^{-1} m_*^{2/3} M_6^{1/3} x_{\rm p}$  is the eccentricity of the most-bound stellar
debris.  The orbital precession of the bound stellar debris leads to the interaction of
the outflowing and inflowing streams at location \citep{dai15}
\begin{equation}
    r_{\rm cr} \simeq {(1+e_{\rm mb}) r_{\rm p} \over 1 - e_{\rm mb} \cos{(\Omega/2)}} \simeq {2 x_{\rm
    p} r_{\rm S}  \over \delta + 2 \sin^2(\Omega/4)} .
    \label{eq:rcros}
\end{equation}
The location of the self-interaction determines the orientation and size of the elliptical accretion disc.

Provided that the inflowing and outflowing steams have similar mass and the collision is completely 
inelastic, the formed elliptical disk has a semimajor axis   \citep{dai15}
\begin{equation}
    a_{\rm d} \simeq {r_{\rm cr} \over 2 \sin^2(\theta_{\rm c}/2)} {1\over   1 + (r_{\rm cr}/ 2 a_{\rm mb})
    \cot^2(\theta_{\rm c}/2)}  ,
    \label{eq:dsize}
\end{equation}
where $a_{\rm mb} \simeq r_{\rm t}^2 /2R_*$ is the orbital semimajor axis of the most-bound stellar
debris nearly independent of the orbital penetration factor $\beta = r_{\rm t}/r_{\rm p}$
\citep{gui13,hay13} and $\theta_{\rm c}$ is the intersection angle of the outgoing stream with the
incoming stream, given by
\begin{equation}
    \cos(\theta_{\rm c}) = {1- 2 \cos(\Omega/2)e_{\rm mb} + \cos(\Omega)e_{\rm mb}^2 \over 1 - 2
    \cos(\Omega/2)
    e_{\rm mb} + e_{\rm mb}^2} ,
\end{equation}
which gives
\begin{equation}
    \sin(\theta_{\rm c}/2) =  {2  \sin(\Omega/4) \cos(\Omega/4)\over  \sqrt{\delta^2 +4\sin^2(\Omega/4) -
    4\delta \sin^2(\Omega/4)} }
     \simeq \cos(\Omega/4)
    \label{eq:intsec}
\end{equation}
\citep{liu17}. From Equation~(\ref{eq:dsize}) together with Equations~(\ref{eq:rcros}) and (\ref{eq:intsec}),
we have
\begin{equation}
a_{\rm d} \simeq  {2  x_{\rm p}   \over 2\delta  +  \sin^2(\Omega/2)} r_{\rm S}
\label{eq:thdisk}
\end{equation}
\citep{liu17}. For low-mass main sequence stars ($0.1  \leq m_* \leq 1$) with $r_* \simeq  m_*^{1-
\zeta}$ and $\zeta\simeq 0.2$ \citep{kip94}, we have $\delta \simeq 8.49\times 10^{-4} m_*^{-2/15} 
M_6^{1/3}  x_{\rm p}$.

In a completely inelastic collision, no angular momentum transfers and redistributes between the
outgoing and incoming streams. The shocked stream has specific angular momentum $j_{\rm
mb} \simeq \sqrt{(1+e_{\rm mb}) r_{\rm p} GM_{\rm BH}}$. The conservation of angular momentum
of the shocked stream gives an eccentricity of the debris accretion disk
\begin{equation}
e_{\rm d} \simeq \left[1-  {(1+e_{\rm mb})r_{\rm p}\over a_{\rm d}}\right]^{1/2} \simeq   \left[\cos^2\left({\Omega
\over 2}\right) - 2 \delta\right]^{1/2}  .
\label{eq:decc}
\end{equation}

\subsection{A  disk model for the broad emission lines}\label{sec:model}

Following \citet{liu17}, we suggest that the broad optical emission lines of TDEs are generated in the elliptical accretion disk presented in Sec.~\ref{ssec:disk}. It is not very clear how the optical emission lines are
excited in the accretion disk. The observations of TDE candidates in the archive of \textit{Swift} Burst Alert
Telescope (BAT) show that there is tentative evidence that hard X-rays may be common in unbeamed 
tidal disruption events \citep{hry16}. A significant fraction of TDEs has been identified and detected
in X-rays \citep[see][for a recent review]{kom15}. The accretion disk in TDEs is ionized and  irradiated by the 
coronal X-ray source. Because the poloidal magnetic field lines are anchored to the ionized accretion
disk and the coronal X-ray source is powered by the  reconnection of the field lines, the coronal 
materials would move nearly radially and become radially extended. 

An ionized accretion disk irradiated by X-ray sources could produce strong optical 
emission lines 
only if  the ionization parameter is low to intermediate \citep{gar13}. To approximate the reflection optical 
line emissivity of the accretion disk irradiated by an extended X-ray source, we adopt the broken power
law in radius $r$ \citep{wil12,gon17} 
\begin{equation}
I_{\nu_{\rm e}} = \left\{
   \begin{array}{r@{\quad}l}
   {\epsilon_0  c \over 2 (2\upi)^{3/2} \sigma} \left({\xi\over \xi_{\rm br}}\right)^{-\alpha_1}
   \exp{\left[-{(\nu_{\rm e}-\nu_0)^2 c^2 \over 2 \nu_0^2 \sigma^2}\right]} & {\rm for} \,
   \xi \leq \xi_{\rm br} \\
   {\epsilon_0  c \over 2 (2\upi)^{3/2} \sigma} \left({\xi\over \xi_{\rm br}}\right)^{-\alpha_2}
   \exp{\left[-{(\nu_{\rm e}-\nu_0)^2 c^2 \over 2 \nu_0^2 \sigma^2}\right]}  & {\rm for} \,
   \xi > \xi_{\rm br}
\end{array} \right.
\label{eq:emiss}
\end{equation}
with $\xi= r/r_{\rm S}$ and  $\xi_{\rm br} =r_{\rm br}/r_{\rm S}$ with $r_{\rm br}$ the radial extent of 
extended X-ray source, where $\nu_{\rm e}$ is the frequency in the frame of the emitter, $\epsilon_0$ 
is a constant to be fitted, $c$ is the speed of light,  $\nu_0$ is the rest frequency of 
the emission line, and $\sigma$ is a velocity dispersion of the local Guassian line broadening because 
of thermal and turbulent motions. In Equation~(\ref{eq:emiss}), the emissivity power index $\alpha_2$ 
is fixed with the typical $\alpha_2\simeq 3$, and $\alpha_1$ depends on the radial distribution of 
X-ray sources and can be used to probe the geometry of the coronal X-ray source \citep{wil12,gon17}.

\begin{figure}
\begin{center}
\includegraphics[width=0.7\columnwidth]{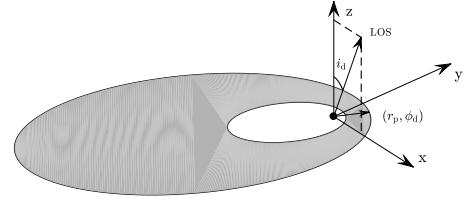}
\caption{General scheme illustrating the geometry and coordinate system used in
the profile calculations. The $z$-axis is along the rotation axis of the accretion disk, and the accretion disk
lies within the $xy$-plane.  The line-of-sight of the observer at infinity is in the $xz$-plane and makes an
inclination angle $i_{\rm d}$ to the $z$-axis.
\label{fig:cor}
}
\end{center}
\end{figure}

The specific flux at frequency $\nu$ received by the observer at infinity is given by
\begin{equation}
f_\nu  =  {r_{\rm S}^2  \cos{i_{\rm d}} \over d^2} \int  \int \xi \,
            \mathrm{d}\xi \, \mathrm{d}\phi \,
            I_{\nu_{\rm e}} D^3(\xi,\phi)\; \psi(\xi,\phi)
\label{metheq:line}
\end{equation}
\citep{era95,def16}, where $\phi$ is the azimuthal angle around the disk with respect to the projected
observer's LOS in the disk plane,  function $\psi(\xi, \phi)$ describes the effects of curved trajectories of 
light rays \citep{bel02,def16}; $D(\xi,\phi) \equiv \nu/\nu_{\rm e}$, and $d$ and $i_{\rm
d}$ are, respectively, the Doppler factor, luminosity distance to the source, and disk inclination angle
with respect to the LOS. The schematic geometry and coordinate systems are shown in Fig.~\ref{fig:cor}. 

The Doppler factor $D(\xi,\phi)$ describes the effects of gravity and the motion of the
emitting particles on the energies of the emitted photons
\begin{eqnarray}
D & = & {(1- 1/\xi)^{1/2} \over \gamma} \bigg\{1- {\beta_{\rm r} [1 - (b/r)^2 (1- 1/\xi)]^{1/2}
      \over 1 - 1/\xi} + \nonumber\\
      && {\beta_{\phi} (b/r) \sin{i_{\rm d}} \sin{\phi} \over (1 - \sin^2{i_{\rm d}}
      \cos^2{\phi})^{1/2}}\bigg\}^{-1}
      \label{eq:doppler}
\end{eqnarray}
\citep{era95}, where $\beta_{\rm r} =  (\mathrm{d}{r}/\mathrm{d}{t})/c $ and $\beta_{\phi} = r
 (\mathrm{d}{\phi}/\mathrm{d}{t})/c$ are, respectively, the radial and
azimuthal velocities of the emitting particles in the source frame, and $\gamma$ is the Lorentz factor
$\gamma = [1 - \beta_{\rm r}^2 (1-1/\xi)^{-2} - \beta_{\phi}^2 (1- 1/\xi)^{-1}]^{-1/2}$.
The ratio ($b/r$) of the impact parameter at infinity and the radius describes how rays emitted from
the debris are mapped to points in the image at infinity \citep{era95,bel02,def16}.

The generalized Newtonian potential in the low-energy limit \citep{tej13} is adopted to describe the 
motion of particles in the disk plane, because it can reproduce exactly the trajectories of particles in
Schwarzschild space-time and the radial dependences of the specific binding energy and
angular momentum of the elliptical orbits. With the generalized Newtonian potential, the radial and 
azimuthal velocities of the emitting particles in the source frame are, respectively, 
\begin{eqnarray}
\beta_{\rm r} &\simeq &   \left( 1 - {1 \over \xi}\right) \sqrt{2 {E_{\rm G} \over
       c^2} + {1 \over \xi} - {l_{\rm G}^2 \over r_{\rm S}^2 c^2 }{1\over \xi^2} \left(1 - {1 \over \xi}\right)} ,\\
  \beta_{\phi} &\simeq &  {l_{\rm G}\over r_{\rm S} c} {\xi - 1
 \over \xi^2} ,
 \label{eq:svel}
\end{eqnarray}
where $l_{\rm G}$  and $E_{\rm G}$ are, respectively, the orbital specific angular momentum and energy
\citep{tej13}. For an elliptical orbit with semimajor axis $a$ and eccentricity $e$, $l_{\rm G}$ and $E_{\rm 
G}$ are fixed, respectively,
\begin{eqnarray}
{l_{\rm G} \over r_{\rm S} c} & = & {(1-e^2) x_{\rm a}  \over \sqrt{2 (1-e^2) x_{\rm a}   - 3 - e^2}} ,
     \label{eq:ang}\\
{E_{\rm G} \over c^2} & = & -{1 \over 2} {x_{\rm a}(1-e^2) - 2 \over x_{\rm a} \left[2 x_{\rm a}(1-e^2) -
	3-e^2\right]} ,
\label{eq:eng}
\end{eqnarray}
where $x_{\rm a}=a/r_{\rm S}$.

It is unclear whether the stream within the accretion disk gets circularized \citep{bon17} or remains highly
eccentric \citep{svi17} at accretion onto the BH. Following \citet{liu17}, we assume that the eccentric disk
consists of nested elliptical annuli with a single eccentricity, which does not vary with semi-major 
axis\footnote{Eccentricity with a power-law distribution in semimajor axis improve the fit little.}, or that the disk 
viscous time is much smaller than the circularization time of stellar debris. The particles in a given elliptical 
annulus of pericenter orientation $\phi_{\rm d}$ have trajectories 
\begin{equation}
r = {a (1 - e^2) \over 1+ e \cos(\phi - \phi_{\rm d})} .
\end{equation}
The disk pericenter orientates toward the observer at $\phi_{\rm d}  = 0^\circ$. 

\subsection{Fitting the H$\alpha$ profiles of ASASSN-14li} \label{spec:fit}

To fit the H$\alpha$ line profiles, we first calculate a set of the model line profiles for the emission-line
region between the inner radius $\xi_1$ with $2\leq \xi_1 \leq 70$ and the outer radius $\xi_2$ with $\xi_2 = 
(1+e_{\rm d}) a_{\rm d}/ r_{\rm S}$. The elliptical accretion disk has an inner edge $2r_{\rm S}$,
because the accretion disk is extremely eccentric ($e_{\rm d} \simeq 0.970$; see Section~\ref{sec:disk}) 
and the disk fluid elements with nearly parabolic orbits passing through the marginal bound orbit $2r_{\rm 
S}$ fall freely on to the BH \citep{abr78}. The disk semimajor axis $a_{\rm d}/r_{\rm S}$ (and thus the disk
outer edge) and eccentricity $e_{\rm d}$ are computed with Equations~(\ref{eq:thdisk}) and (\ref{eq:decc}) 
for the star's orbital pericenter $x_{\rm p}$ for $1 \leq x_{\rm p} \leq 50$. In modeling the line profiles, the 
BH mass is required only in the calculation of the disk semimajor axis and eccentricity through the small 
quantity $\delta$ in Equations~(\ref{eq:thdisk}) and (\ref{eq:decc}).  We do not estimate the BH mass 
by modeling the line profiles and adopted the BH mass $\lg{\left(M_{\rm BH}/{\rm M}_\odot\right)} = 
6.23^{+0.39}_{-0.40}$ obtained with the $M_{\rm BH} - \sigma_*$ correlation of the host bulge and central 
BH mass \citep{wev17}, because the model line profiles are mainly determined by both the kinematic 
Doppler and gravitational lensing effects, which are functions of dimensionless radius $\xi$ instead of 
the dimensional radius $r$ (c.f. Equations~\ref{eq:doppler} and \ref{eq:svel}). Having the BH mass 
as a free parameter would improve fits little and the estimate of BH mass has much larger uncertainties 
than that obtained from the $M_{\rm BH} - \sigma_*$ relation. 

The model line profiles are also computed for the parameter space, $0 \leq i_{\rm d} \leq \upi$, 
$0\leq \phi_{\rm d} < 2\upi$, $\xi_{\rm br} \leq \xi_2$, $5\times 10^{-2} \times {\rm FWHM} \leq 
\sigma \leq (2/3) \times {\rm FWHM}$ and $0\leq \alpha_1 \leq 3$. The observed H$\alpha$ line 
profiles of the 12 spectra are jointly fitted with shared periapse $x_{\rm p}$, inner radius $\xi_1$, broken 
radius radius $\xi_{\rm br}$, emissivity power-law index $\alpha_1$, and velocity dispersion $\sigma$ 
with the least-square method ($\chi^2$).

\section{Results}
\label{sec:disk}

\begin{figure*}
\begin{center}
\includegraphics[width=1.8\columnwidth]{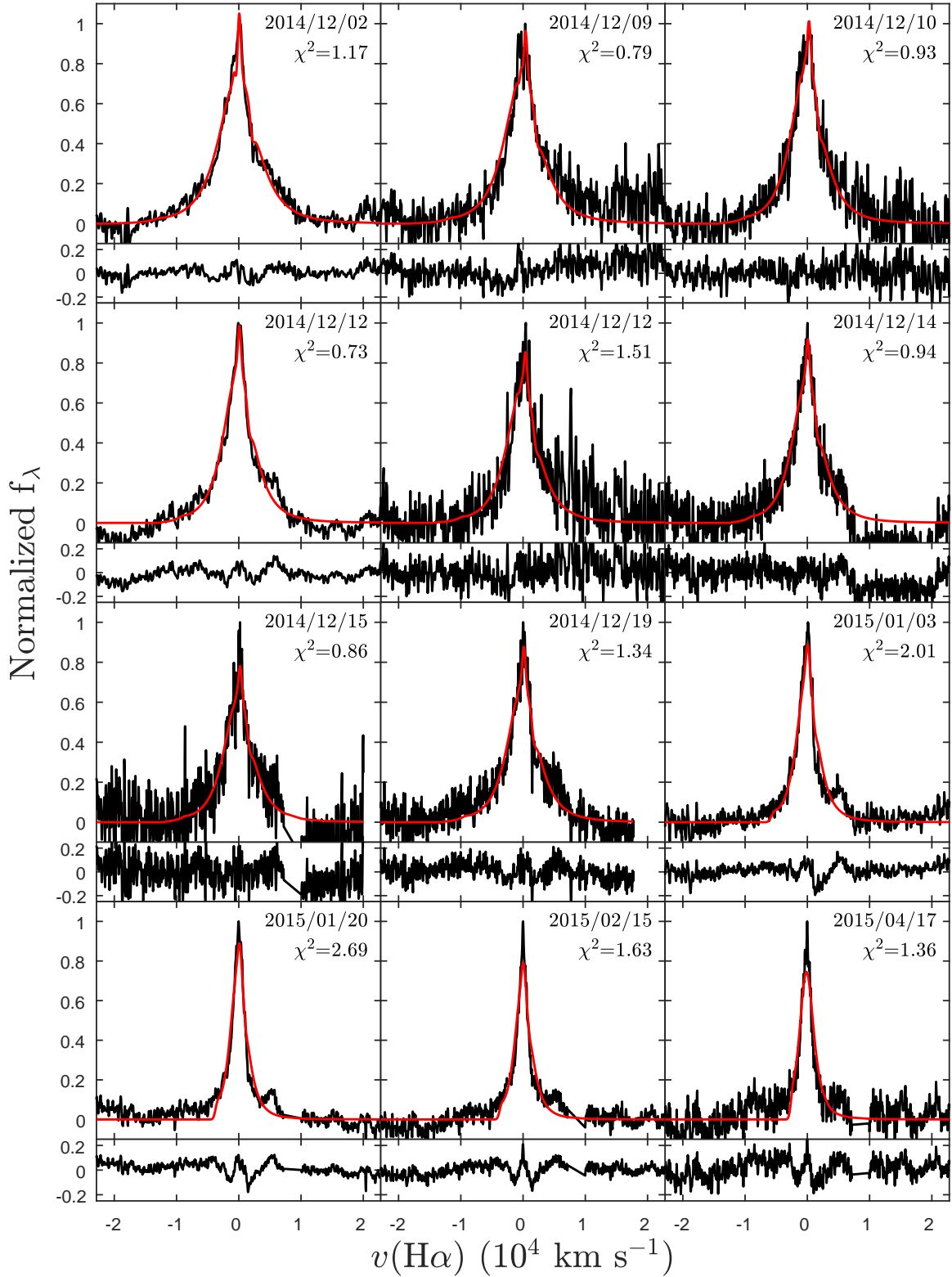}
\caption{Normalized velocity profiles of H$\alpha$ and fits of the model spectra for different epochs. 
The UT date (year/month/day) is given at the top right of each panel.  The observed profiles (black) 
are complex and asymmetric but are well reproduced with the relativistic elliptical disk model  (red 
solid). The residuals of the spectra, after subtraction of the fit, are given at the bottom of each panel. 
The reduced $\chi^2$ of the best fit is given at the top right of each panel.
\label{fig:halpha}
}
\end{center}
\end{figure*}

Fig.~\ref{fig:halpha} shows the best-fit models of the 12 H$\alpha$ spectra and the reduced $\chi^2$. 
The reduced $\chi^2$ is calculated with respect to the averaged noise level over the closest regions of 
the emission lines after subtraction of the spectral features. Table~\ref{tab:diskpar} gives the best-fit
values of the model parameters and their associated uncertainties at 90\% confidence level. We calculated 
the uncertainties of the fitting parameters at 90\% confidence level with the Markov Chain Monte Carlo 
(MCMC) method \citep{liu17}

The results indicate that the broad complex and asymmetric emission-line profiles of the 12 H$\alpha$ spectra 
of TDE ASASSN-14li can be well fitted with a relativistic elliptical accretion disk model. The elliptical accretion
disk has an extreme eccentricity $e_{\rm d} \simeq 0.9700^{+0.0002}_{-0.0013}$ and large semimajor 
axis $a_{\rm d} \simeq 847.0^{+23.3}_{-5.3} r_{\rm S}$ (or apogee $1668.5 r_{\rm S}$). The asymmetric
and complex substructures of the line profiles are mainly due to the elliptical motion of the line-emitting matter 
within the disk plane with low inclination $i_{\rm d} \sim 26\degr$ and pericenter orientation 
$\phi_{\rm d}\simeq 2.7\degr$, closely toward the observer. The elliptical accretion disk forms 
following the tidal disruption of a star with orbital pericenter $r_{\rm p} \simeq 25.33^{+1.82}_{-0.36}
 r_{\rm S}$, passing by a BH of mass $M_{\rm BH} \simeq 10^{6} {\rm M_\odot}$ with penetration factor 
$\beta = r_{\rm t}/r_{\rm p} \simeq 0.93 m_*^{-\zeta +2/3} M_6^{-2/3}$.

Our results show that the broad optical emission lines originate from the regions of the elliptical
accretion disk from about $9r_{\rm S}$ to the disk apogee $1668.5 r_{\rm S}$ and is locally 
broadened with Gaussian velocity dispersion $\sigma\simeq 279 \, {\rm km\; s^{-1}}$,  about 17 
times smaller than the local broadening of the H$\alpha$ line in the TDE candidate PTF09djl \citep{liu17}. 
The elliptical accretion disk is overlapped by an extended hot corona of radial extent $r_{\rm 
br} \simeq 1668.5 r_{\rm S}$, or the entire disk radial extent. The disk is illuminated by the extended 
coronal X-ray source, and the line emissivity of disk reflection spectra is practically a single power-law 
in radius with index $\alpha_1 \simeq -0.049^{+0.035}_{-0.032}$ for the entire disk region covered by 
the corona. A power-law index $\alpha_1 \simeq 0$ is expected with an extended illuminating coronal 
X-ray source of flat radial distribution \citep{wil12,gon17}.

\begin{figure}
\begin{center}
\includegraphics[width=0.9\columnwidth]{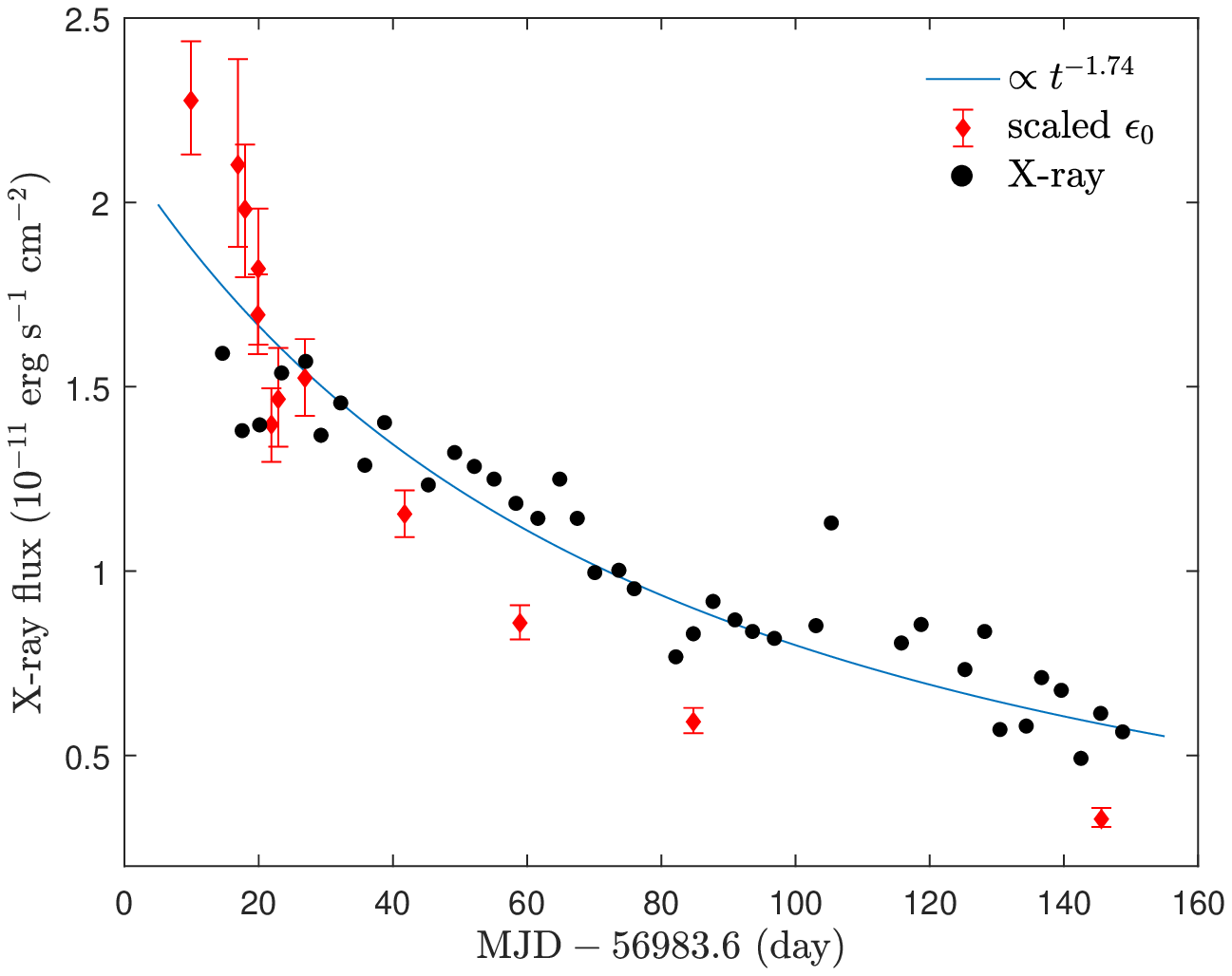}
\caption{Evolution of the line emissivity constant $\epsilon_0$ (diamonds) over 
time, starting at discovery (MJD = 56983.6 days). The filled circles are the 0.3-10 keV X-ray 
observations \citep{mil15}. The solid line gives a power-law fit of the X-ray flux. The line emissivity 
constant  $\epsilon_0$ is scaled with a constant $1.08\times10^{-40}$~sr~\AA~and closely 
follows the X-rays at 0.3-10 keV, 
implying that the optical emission lines in TDEs are powered by the X-ray source.
\label{fig:epsilon}}
\end{center}
\end{figure}

Fig.~\ref{fig:epsilon} gives for comparison the evolution of the best-fit line emissivity constant $\epsilon_0$ 
over time and the X-ray flux at 0.3-10 keV. The data of the {\it Swift} X-ray flux at 0.3-10 keV for the 
period of the observations of the 12 optical spectra are taken from \citet{mil15}. The evolution of 
the X-ray flux is well fitted by a power-law $\propto t^{-1.74}$ and $\epsilon_0$ is scaled by a 
constant $1.08\times10^{-40}$~sr~\AA~to match the X-ray observations. Fig.~\ref{fig:epsilon} 
shows that the line emissivity constant $\epsilon_0$ roughly follows the X-ray flux at 0.3-10 keV, 
implying a common energy source of the optical emission lines and the X-ray emission.

\begin{figure}
\begin{center}
\includegraphics[width=0.9\columnwidth]{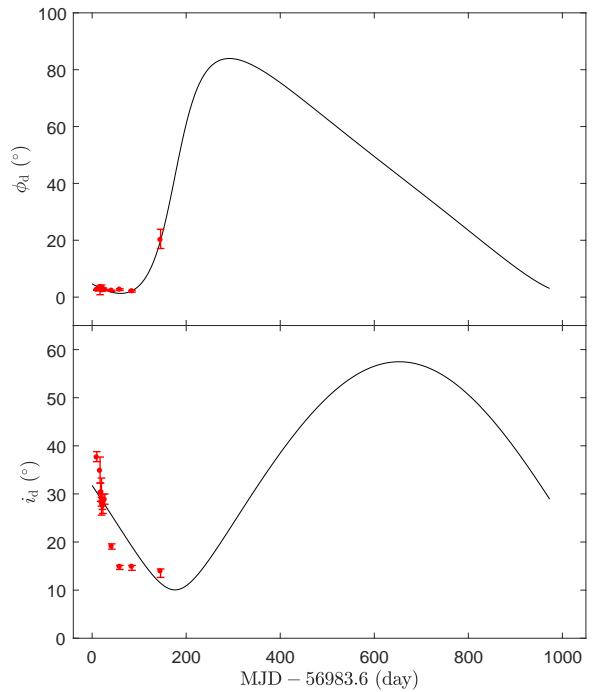}
\caption{Disk periastron azimuthal angle $\phi_{\rm d}$ (upper) and inclination $i_{\rm d}$ (lower)
versus observational time, starting at discovery (MJD = 56983.6 days). The azimuthal angle is nearly
constant for most time of the spectral observations and the inclination decreases with time, which are 
consistent with the precession of an elliptical accretion disk (solid) because of the Lense-Thirring effect 
with precession period $T_{\rm LS} \simeq 953.7\, {\rm days}$. The associated uncertainties at 
90\% confidence level calculated with MCMC are plotted.
\label{fig:incl}}
\end{center}
\end{figure}

\begin{figure}
\begin{center}
\includegraphics[width=0.9\columnwidth]{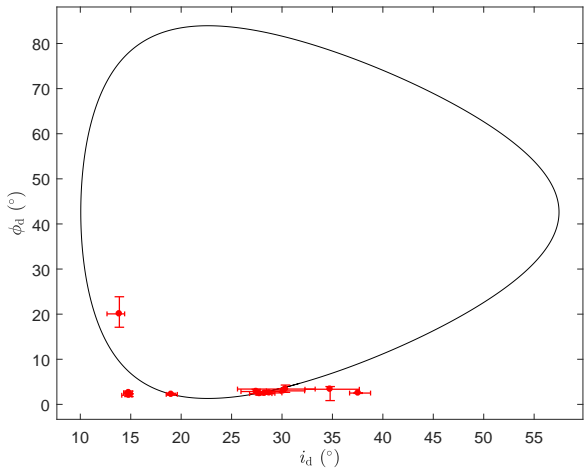}
\caption{The relation between the disk azimuthal angle $\phi_{\rm d}$ and inclination $i_{\rm d}$
(filled circle),  consistent with the clockwise precession of an elliptical accretion disk (solid). The
rotation axis of the elliptical accretion disk misaligns by about $23.7\degr$ with the spin axis of BH, which
inclines with respect to the LOS by about $33.8\degr$. The associated uncertainties at 90\% confidence
level calculated with MCMC are plotted.
\label{fig:iphi}
}
\end{center}
\end{figure}

Fig.~\ref{fig:incl} gives the disk pericenter  position and inclination angles as a function of the
observational time.  The disk pericenter position angle $\phi_{\rm d}$ is nearly constant with 
$\phi_{\rm d}\simeq 2.7\degr$ except that of the last spectra with $\phi_{\rm d}\simeq 20\degr$, 
but the disk inclination decreases from about $38\degr$ at early epochs to $14\degr$ about 146 days 
after discovery. The changes in the line profiles are mainly due to the variations of the disk inclination.
Because the orientation of the disk
semimajor axis is determined by the location of the intersection of the outflowing and inflowing
streams, a constant disk semimajor axis would imply an invariant orientation of the disk semimajor.
However, the precession of elliptical disk would lead to an associated variation of the disk azimuthal
angle. Fig.~\ref{fig:incl} shows that the observations of the disk inclination $i_{\rm d}$ and pericenter
azimuthal angles $\phi_{\rm d}$ are consistent with the precession of the elliptical accretion disk with
precession period $T_{\rm LS} \simeq 954\,{\rm days}$. The $\phi_{\rm d}$--$i_{\rm d}$ 
diagram in Fig.~\ref{fig:iphi} shows that the relationship of the disk azimuthal angle $\phi_{\rm d}$ 
and inclination $i_{\rm d}$ is consistent with the expectation of the disk precession. The elliptical accretion 
disk precesses due to the Lense-Thirring effect because of the misalignment of the disk rotation axis 
and BH spin by an angle about $23.7\degr$. The BH spin axis inclines with respect to the LOS by 
about $33.8\degr$.  Because the spectral observational campaign lasted only about 14\%  
of the disk precession period $T_{\rm TS}$, the present observations cannot give useful constraints
on the parameters of disk precession, and much more spectral observations of high resolutions and
signal-to-noise ratios are needed.


\begin{table*}
\small
\caption{Best-fitting disk parameters and associated uncertainties for the broad H$\alpha$ emission line.
\label{tab:diskpar}}
\begin{tabular}{ccccccccccc}
\hline
 Date&$r_{\rm p}$ & $\phi_{\rm d}$  & $i_{\rm d}$  & $a_{\rm d}$ & $e_{\rm d}$ &$r_1$&$r_{\rm br}$
 & $\alpha_1$ &$\sigma$  &$\epsilon_0$ \\
 & ($r_{\rm S}$) &  ($^\circ $) &  ($^\circ$) &  ($r_{\rm S}$) &  & ($r_{\rm S}$) & ($r_{\rm S}$)&  &($\rm km \; s^{-1}$) & \\
\hline
\hline
2014/12/02 & $25.33^{+1.82}_{-0.36}$&2.5$^{+0.1}_{-0.1}$&37.6$^{+1.2}_{-0.9}$&847.0$^{+23.3}_{-5.3}$&
0.9700$^{+0.0002}_{-0.0013}$&8.96$^{+1.82}_{-2.12}$&1668.5$^{+44.8}_{-10.2}$&$-0.049^{+0.035}_{-0.032}$&279$^{+11}_{-10}$&$21.1^{+1.5}_{-1.4}$\\
2014/12/09 & $25.33^{+1.82}_{-0.36}$&3.4$^{+0.6}_{-2.5}$&34.8$^{+2.9}_{-2.5}$&847.0$^{+23.3}_{-5.3}$&
0.9700$^{+0.0002}_{-0.0013}$&8.96$^{+1.82}_{-2.12}$&1668.5$^{+44.8}_{-10.2}$&$-0.049^{+0.035}_{-0.032}$&279$^{+11}_{-10}$&$19.5^{+2.7}_{-2.1}$\\
2014/12/10 & $25.33^{+1.82}_{-0.36}$&3.0$^{+0.5}_{-0.3}$&30.1$^{+2.1}_{-1.7}$&847.0$^{+23.3}_{-5.3}$&
0.9700$^{+0.0002}_{-0.0013}$&8.96$^{+1.82}_{-2.12}$&1668.5$^{+44.8}_{-10.2}$&$-0.049^{+0.035}_{-0.032}$&279$^{+11}_{-10}$&$18.4^{+1.6}_{-1.7}$\\
2014/12/12 & $25.33^{+1.82}_{-0.36}$&2.5$^{+0.2}_{-0.2}$&28.3$^{+1.0}_{-0.8}$&847.0$^{+23.3}_{-5.3}$&
0.9700$^{+0.0002}_{-0.0013}$&8.96$^{+1.82}_{-2.12}$&1668.5$^{+44.8}_{-10.2}$&$-0.049^{+0.035}_{-0.032}$&279$^{+11}_{-10}$&$15.7^{+1.0}_{-1.0}$\\
2014/12/12 & $25.33^{+1.82}_{-0.36}$&3.4$^{+0.9}_{-0.7}$&30.3$^{+2.9}_{-4.8}$&847.0$^{+23.3}_{-5.3}$&
0.9700$^{+0.0002}_{-0.0013}$&8.96$^{+1.82}_{-2.12}$&1668.5$^{+44.8}_{-10.2}$&$-0.049^{+0.035}_{-0.032}$&279$^{+11}_{-10}$&$16.9^{+1.5}_{-1.9}$\\
2014/12/14 & $25.33^{+1.82}_{-0.36}$&2.4$^{+0.4}_{-0.3}$&27.8$^{+1.2}_{-1.0}$&847.0$^{+23.3}_{-5.3}$&
0.9700$^{+0.0002}_{-0.0013}$&8.96$^{+1.82}_{-2.12}$&1668.5$^{+44.8}_{-10.2}$&$-0.049^{+0.035}_{-0.032}$&279$^{+11}_{-10}$&$12.9^{+0.9}_{-0.9}$\\
2014/12/15 & $25.33^{+1.82}_{-0.36}$&2.8$^{+0.5}_{-0.4}$&27.5$^{+2.4}_{-1.5}$&847.0$^{+23.3}_{-5.3}$&
0.9700$^{+0.0002}_{-0.0013}$&8.96$^{+1.82}_{-2.12}$&1668.5$^{+44.8}_{-10.2}$&$-0.049^{+0.035}_{-0.032}$&279$^{+11}_{-10}$&$13.6^{+1.3}_{-1.2}$\\
2014/12/19 & $25.33^{+1.82}_{-0.36}$&2.5$^{+0.2}_{-0.2}$&28.7$^{+1.2}_{-0.9}$&847.0$^{+23.3}_{-5.3}$&
0.9700$^{+0.0002}_{-0.0013}$&8.96$^{+1.82}_{-2.12}$&1668.5$^{+44.8}_{-10.2}$&$-0.049^{+0.035}_{-0.032}$&279$^{+11}_{-10}$&$14.1^{+1.0}_{-1.0}$\\
2015/01/03 & $25.33^{+1.82}_{-0.36}$&2.2$^{+0.2}_{-0.2}$&19.0$^{+0.6}_{-0.5}$&847.0$^{+23.3}_{-5.3}$&
0.9700$^{+0.0002}_{-0.0013}$&8.96$^{+1.82}_{-2.12}$&1668.5$^{+44.8}_{-10.2}$&$-0.049^{+0.035}_{-0.032}$&279$^{+11}_{-10}$&$10.7^{+0.6}_{-0.6}$\\
2015/01/20 & $25.33^{+1.82}_{-0.36}$&2.6$^{+0.3}_{-0.2}$&14.8$^{+0.3}_{-0.5}$&847.0$^{+23.3}_{-5.3}$&
0.9700$^{+0.0002}_{-0.0013}$&8.96$^{+1.82}_{-2.12}$&1668.5$^{+44.8}_{-10.2}$&$-0.049^{+0.035}_{-0.032}$&279$^{+11}_{-10}$&$8.0^{+0.4}_{-0.4}$\\
2015/02/15 & $25.33^{+1.82}_{-0.36}$&2.0$^{+0.3}_{-0.3}$&14.8$^{+0.3}_{-0.7}$&847.0$^{+23.3}_{-5.3}$&
0.9700$^{+0.0002}_{-0.0013}$&8.96$^{+1.82}_{-2.12}$&1668.5$^{+44.8}_{-10.2}$&$-0.049^{+0.035}_{-0.032}$&279$^{+11}_{-10}$&$5.5^{+0.3}_{-0.3}$\\
2015/04/17 &  $25.33^{+1.82}_{-0.36}$&20.1$^{+3.8}_{-3.0}$&13.9$^{+0.5}_{-1.2}$&847.0$^{+23.3}_{-5.3}$&
0.9700$^{+0.0002}_{-0.0013}$&8.96$^{+1.82}_{-2.12}$&1668.5$^{+44.8}_{-10.2}$&$-0.049^{+0.035}_{-0.032}$&279$^{+11}_{-10}$&$3.0^{+0.3}_{-0.2}$\\
\hline
\end{tabular}
\begin{flushleft}
{\it Note}. -- The associated uncertainties at 90\% confidence level
are calculated with the MCMC method. Cols (1)-(11):  Date: observational time (year/month/day);
$r_{\rm p}$: orbital pericenter of star; $\phi_{\rm d}$: disk-pericenter orientation with respect to the
LOS projected in the disk plane; $i_{\rm d}$: inclination angle of the disk rotation axis with respect to
the LOS; $a_{\rm d}$: disk semimajor; $e_{\rm d}$: disk eccentricity; $r_1$($r_1 = \xi_1 r_{\rm S}$): inner
radius of the
emission line regions; $r_{\rm br}$: radial extent of the coronal X-ray sources; $\alpha_1$: emissivity
power-law index of inner disk regions; $\sigma$: local velocity dispersion of emission line.; $\epsilon_0$
($10^{28}$~erg~s$^{-1}$~cm$^{-2}$~\AA$^{-1}$~sr$^{-1}$): constant of the reflection optical line 
emissivity of the accretion disk.
\end{flushleft}
\end{table*}

\section{Discussion and conclusion}
\label{sec:dis}

We have analyzed a set of 12 optical spectra of the TDE  ASASSN-14li, after careful treatment of
the TDE featureless continuum and host galaxy starlight. We have successfully modelled the
complex, asymmetric and single-peaked substructures of the broad H$\alpha$ emission line
with a relativistic elliptical accretion disk model, although the profiles of the optical emission lines
of ASASSN-14li are  radically different from those of the double-peaked broad H$\alpha$ line
of the TDE candidate PTF09djl \citep{liu17}.  The results  show that the TDE ASASSN-14li is powered
by the tidal disruption of a star,  passing by a BH of mass $M_{\rm BH} \simeq 10^{6.23} M_\odot$
with an orbital pericenter $r_{\rm p}\simeq 25.33 r_{\rm S}$ or penetration factor $\beta 
\simeq 0.93 m_*^{-\zeta +2/3} M_6^{-2/3}$. The returned bound stellar debris forms an elliptical 
accretion disk of a semimajor $a_{\rm d}\simeq 847.0 r_{\rm S}$ and eccentricity $e_{\rm 
d}\simeq 0.970$. The elliptical accretion disk is irradiated  by an extended coronal X-ray source  
with radial size of the total radial extent of the elliptical accretion disk. The coronal X-ray
source has a flat radial radial distribution with a power-law of index $\alpha_1\simeq 
-0.049$. The debris accretion disk of ASASSN-14li has a nearly constant pericenter azimuthal 
position angle with $\phi_{\rm d} \simeq 2.7\degr$ for most of the spectral observational time 
except that for the last spectral observation and decreasing inclination from $i_{\rm d} \simeq 
38\degr$ at early epoch to $14\degr$ at the last spectral observation.

The pericenter azimuthal position angle $\phi_{\rm d} \sim 2.7\degr$ suggests that the 
major axis of the elliptical accretion disk of ASASSN-14li nearly points toward the observer. 
The major axis of the elliptical accretion disk of the TDE candidate PTF09djl has a large disk pericenter 
position angle $\phi_{\rm d} \simeq 72\degr$ \citep{liu17}. The orientations of the major axis of elliptical 
accretion disk in the two TDEs have a relative position angle $\Delta{\phi}_{\rm d} \simeq 
69\degr$ and are consistent with a roughly uniformly distribution of $\phi_{\rm d}$ between $0\degr$ 
and $360\degr$ or of $\Delta{\phi}_{\rm d}$ between $0\degr$ and $90\degr$. We will check the 
conclusion by modeling the broad optical emission lines for a larger sample of TDEs. 

The orientations of the elliptical accretion disk of the two TDEs are very different
with respect to the observer and the profiles of optical emission lines strongly depend on the position 
angle $\phi_{\rm d}$ for extremely eccentric disks. The profiles of the optical emission lines of the 
TDEs ASASSN-14li and PTF09djl are significantly different, being moderately broad and single-peaked 
for the former and extremely broad and double-peaked
for the latter. For a power-law disk emissivity $I_{\nu_{\rm e}} \sim \xi^{-\alpha}$, the line flux from the disk 
region at radius $r$ is about $f_\nu\sim \xi^{2-\alpha}$. For a lamppost or compact illuminating X-ray 
source of  power-law index $\alpha \sim 3$, as for PTF09djl \citep{liu17}, the line flux is mainly from 
the disk region at small radius $\xi$ and would be significantly Doppler shifted by the large velocity of 
emitting particles. The optical lines may have double-peaked and Doppler-shifted profiles for moderate 
to large disk inclination, depending on the disk pericenter orientations $\phi_{\rm d}$. In contrast,
for an extended flat corona with power-law index $\alpha \simeq -0.049$, as for ASASSN-14li, the 
line emission is dominated by the flux from the disk surface at large radius $\xi$, where the kinematic 
velocity of the emitting particles is relatively small and the Doppler effect is insignificant. It would 
be expected that the optical emission lines are single-peaked with broad and asymmetric wings and 
the peak of line profiles may be low to moderately Doppler-shifted, depending on the disk pericenter
orientation $\phi_{\rm d}$ and inclination $i_{\rm d}$. The observation that the optical emission lines in 
most optical/UV TDEs have asymmetric and single-peaked profiles may suggest that the optical emission 
lines in the TDEs are powered by an extended X-ray source with power-law index $\alpha \la 2$.

The time dependence of the disk inclination and pericenter azimuthal position angles is consistent
with the precession of the elliptical accretion disk with period $T_{\rm LS} \simeq 954 \, {\rm days}$,
about 7 times longer than the duration of the spectral observational campaign. The accretion disk
in ASASSN-14li is nearly face-on with an inclination angle $i_{\rm d}$ from about $38\degr$ to 
$14\degr$ with respect to the LOS, which is much smaller than the inclination angle $i_{\rm d}\simeq 
88\degr$ in the TDE candidate PTF09djl \citep{liu17}. The viewing angle effects must be significant in 
PTF09djl but small in ASASSN-14li. Because the He emission lines are produced in a region of ionized
accretion disk with electron scattering optical depth ($\tau_{\rm es}$) a few times larger than that for 
H lines \citep{rot16}, He emission lines are expected to be strong in the optical spectra of TDEs with 
face-on accretion disk. Because the effective optical depth $\tau_{\rm eff}$ changes with disc inclination
angle, $\tau_{\rm eff} = \tau_{\rm es} /\cos(i_{\rm d})$, the escaped line photons are from shallower 
layers with the increase of disk inclination angle, and consequently both the He and H emission lines are 
attenuated. However, He line photons are produced in deeper layers with larger $\tau_{\rm es}$
and higher temperature than that of H line photons, so the attenuation of He lines is much more 
significant. The viewing angle effect of the large disk inclination $i_{\rm d}\simeq 88\degr$
would lead to the weakness/absence of He emission lines in the optical spectra of the TDE candidate
PTF09djl, while the disk inclination angle $i_{\rm d}$ decreasing from about  $38\degr$ 
to about $14\degr$ is consistent with the prominent detection of the He emission 
lines with the increasing enhanced intensities relative to the H emission lines in the optical spectra of 
ASASSN-14li, as reported by \citet{hun17}. This is consistent with the scenario suggested by \citet{liu17} 
that the diversities of the relative intensity of He and H emission lines in the spectra of the optical/UV TDEs 
and candidates are due to different disk inclinations.

For an eccentricity $e=0.9700$, Equation~(\ref{eq:eng}) gives a conversion efficiency of $\eta \simeq
3.69\times 10^{-3}$ before the stellar debris passing the marginal stable orbit about $2r_{\rm S}$ falls freely 
onto the BH, consistent with the suggestion of low efficiency in TDEs \citep{svi17}. For a stellar tidal 
disruption by SMBH of mass $M_{\rm BH} \simeq 10^{6.23} M_\odot$ with
$\beta\sim 1$, the fallback stellar debris mass rate has a peak $\dot{M}_{\rm p} \sim A_{5/3}
(M_{\rm BH}/10^6 M_\odot)^{-1/2} {\rm M_\odot \; yr^{-1}} \sim 1.02 {\rm M_\odot \; yr^{-1}}$
with $A_{5/3} \simeq 1.33$ \citep{gui13}, which gives a total peak luminosity $L_{\rm p} \simeq
\eta  \dot{M}_{\rm p} c^2 \sim 2.1\times 10^{44}\, {\rm erg \; s^{-1}}$.
Because the Eddington luminosity for mass
$M_{\rm BH} = 10^{6.23} M_\odot$ is $L_{\rm Edd} \simeq 2.1 \times 10^{44} \, {\rm erg\; s^{-1}}$,
we have $L_{\rm p} \sim  L_{\rm Edd}$. The peak accretion of the TDE ASASSN-14li
radiates at about the Eddington limit. ASASSN-14li has been detected in X-rays
\citep{mil15,hol16a,van16,kar17},
optical/UV \citep{hol16a,bro17}, infrared \citep{jia16}, and radio \citep{van16,ale16}.  The integrated
peak optical/UV and X-ray luminosities are $L_{\rm opt}\sim 6.1\times 10^{43} \, {\rm erg\; s^{-1}}$
\citep{hol16a} and $L_{\rm X}\sim 3\times 10^{43} \, {\rm erg\; s^{-1}}$ in the range 0.3--10~keV
\citep{mil15,hol16a,van16,kar17},
respectively. The energy output of ASASSN-14li is dominated in the X-ray and optical/UV bands
and has a peak bolometric luminosity $L_{\rm bol} \sim L_{\rm opt} + L_{\rm X} \sim 0.9\times
10^{44} \, {\rm erg\; s^{-1}}$,  consistent with the expected maximum luminosity $L_{\rm p}$.

It is not very clear how TDEs excite the strong optical and UV emission lines. In the accretion disk model 
for the broad optical emission lines, there are three possible energy sources: (1) a radially extended 
hot corona above the accretion disk surfaces, (2) the interaction shocks at nearly
the apogee of the accretion disk \citep{pir15}, and (3) the compressing shocks at the pericenter of the
accretion disk. We have modeled the line emissivity with a broken power-law in radius. If the shocks at
about the apogee of the accretion disk are the driving sources of emission lines, the break radius $r_{\rm
br}$ should be about the locations of the shocks at near the disk apogee and the line emissivity $\alpha_1$
is negative \citep{wil12,gon17}.  The compressing shocks at the disk pericenter may heat the accretion
disk to radiate soft X-rays \citep{pir15,kro16}, and the hot materials may emit optical emission lines by
recombinations when they move out radially and cool down. If the compressing shocks at the disk
pericenter are the driving energy sources of the emission lines, the line emissivity would have a single
power-law in radius, and the radius $r_{\rm  br}$ would be at about the disk radius at which the disk
central temperature drops below the Hydrogen recombination temperature. The emission lines
are most probably powered by extended coronal X-ray sources, and the broken radius $r_{\rm br}$ is
the radial extent of the coronal X-ray sources. The coronal X-ray source is powered by local dissipation 
of magnetic flux due to the reconnection of the magnetic field lines anchoring the ionized accretion 
disk underneath. In circular accretion disk, the toroidal magnetic fields may be generated by 
magneto-rotational instability (MRI) and are amplified by differential rotation. The magnetic flux is 
proportional to the local orbital shear $\Omega$ with $\Omega$ the orbital angular frequencies. 
The magnetic field rises out of the disk surface due to Parker instability and dissipates locally because 
of the reconnection of magnetic field loops \citep{fie93,par03,par14,beg15}. The MRI turbulence 
regenerates the magnetic fields on a dynamical timescale \citep{bal91}. The dissipation of magnetic 
flux powers the emission lines, probably leading to a power-law radial distribution of H$\alpha$ line
emissivity in stars and cataclysmic variables with circular accretion disk \citep{hor91}. It is unclear 
what kind of radial configuration the magnetic flux has in an extremely eccentric accretion disk. The 
power-law index $\alpha_1 = -0.049$ in the TDE ASASSN-14li is consistent with a flat coronal X-ray source.

For an ionized optically thick accretion disk irradiated by X-rays, the reflected spectra have prominent 
optical emission lines, when the ionization parameter is in the range $1 \leq \zeta {\rm (erg \; cm \; s^{-1})}
\la 500$ \citep{gar13}, where $\zeta = 4\upi F_{\rm x} / n_{\rm e}$ with $F_{\rm x}$ the integrated flux in 
the energy range $0.1-300\, {\rm keV}$ and $n_{\rm e}$ the electron number density. From 
Equation~(\ref{eq:eng}), the accretion disk of semimajor $a_{\rm d} \simeq 847.0 r_{\rm S}$ and eccentricity 
$e_{\rm d}\simeq 0.970$ has a typical peak temperature $T_{\rm p} \simeq 3 \times 10^4 
\, {\rm K}$. The elliptical accretion disc of ASASSN-14li is ionized. The X-ray spectra of ASASSN-14li are soft
\citep{mil15,hol16a,van16,kar17}. Provided that about ten percent the X-ray luminosity $L_{\rm x}
\simeq 3\times 10^{43}  \, {\rm ergs \; s^{-1}}$ in the 0.3--10 keV range \citep{hol16a} or one percent
of the extrapolated X-ray luminosity $L_{\rm x} \simeq 3.2\times 10^{44}  \, {\rm ergs \; s^{-1}}$ in
the 0.1--10 keV range \citep{mil15} is  from an extended coronal X-ray source, the ionization parameter
of ASASSN-14li is $\zeta \sim 6\upi L_{\rm x} (m_{\rm p}/1.2M_*) (1+e_{\rm d}) a_{\rm d} (H/r) [1+
(H/r)^2]^{-1} \simeq 3.3 \, {\rm erg \; cm \; s^{-1}}$ for $M_* \sim M_\odot$ and disk opening angle
$H/r\sim 0.1$, where $m_{\rm p}$ is the mass of proton and the typical disk mass at peak $M_{\rm
d} \sim M_*/3$ is assumed.

The fits to the weak wings of the emission lines can be slightly improved if a bulk velocity $v_{\rm 
m} \sim - 1100 \, {\rm km \; s^{-1}}$ is included in the modeling. Because the wings of the broad emission
lines are noisy, the constraint on the bulk velocity is very weak. Recent investigations show that 
the thickness of a circular accretion disk can modify the wings and shift the peak of the line profiles 
due to the self-shadowing of the accretion disk and to the irradiation of the central 'eye wall' of the 
inner disk \citep{tay18}. The fitting to the wings and peak of the line profiles with a thin accretion disk 
model would result in some biases in estimates of the parameters of the system, in particular the disk 
position angle $\phi_{\rm d}$ and inclination $i_{\rm d}$. It is unclear what are the effects of the height 
of an extremely eccentric accretion disk on the model line profiles. To fully understand the effects 
of the disk height on the line profiles and estimates of the disk parameters,  a more elaborated elliptical
accretion disk model and observed spectra of high signal-to-noise ratio are needed.

In conclusion, we have successfully modeled the complex and asymmetric H$\alpha$ profiles of the
TDE ASASSN-14li with a relativistic elliptical disk model. The accretion disk in ASASSN-14li has a large
semimajor axis and extreme eccentricity and is probably covered by a largely extended coronal 
X-ray source of power-law radial distribution.  The coronal X-ray sources in TDEs have 
very different structures from those in active
galactic nuclei (AGNs) or Galactic X-ray binaries, which are compact.  Our results show that modelling
the complex optical line profiles is powerful in probing the structures of the accretion disk and coronal
X-ray sources.


\section*{Acknowledgements}

We are grateful to Jose Luis Prieto and Subo Dong for kindly providing us the electronic data of the
spectra.
This work is supported by the National Natural Science Foundation of China (NSFC11473003) and the
Strategic Priority Research Program of the Chinese Academy of Sciences (Grant No. XDB23010200 and
No. XDB23040000). LCH was supported by the National Key R\&D Program of China (2016YFA0400702)
and the National Science Foundation of China (11473002, 11721303).

\bsp	
\label{lastpage}
\end{document}